\documentstyle[epsf,sprocl]{article}

\input{psfig}
\def\apj{{\em Ap. J.}}
\def\mnras{{\em M.N.R.A.S.}}
\def\aa{{\em A \& A}}
\def\etal{{\em et al.~}}
\def\lesssim{\mathrel{\hbox{\rlap{\hbox{\lower4pt\hbox{$\sim$}}}\hbox{$<$}}}}
\def\gtrsim{\mathrel{\hbox{\rlap{\hbox{\lower4pt\hbox{$\sim$}}}\hbox{$>$}}}}

\def\mco{\multicolumn}

\def\ra{\rightarrow}

\def\ko{K^0}

\def\be{\begin{equation}}
\def\ee{\end{equation}}
\def\bea{\begin{eqnarray}}
\def\eea{\end{eqnarray}}

\def\thefirstfig{
\makebox{
\medskip
\noindent
\parbox[l]{1.7truein}{
\footnotesize
{\bf Figure 1. CMB Data}\\
A compilation of 24 of the most recent
measurements of the CMB angular power spectrum.
Models with $h=0.30$ and $h=0.75$ are superimposed
(both are $\Omega = 1$, $\Omega_{b}=0.05$,  $n=1$ $Q=18\;\mu$K models).
The dotted line is a 5th order polynomial fit to the data.
The low-$h$ value is favored.
MAP and Planck Surveyor are expected to yield precise spectra for 
$\theta_{FWHM} \gtrsim 0^{\circ}\!.3$ and 
$\theta_{FWHM} \gtrsim 0^{\circ}\!.2$ respectively
(see angular scale marked at the top).
Figure from Lineweaver \etal (1997a).
}
\hglue0.1truein
\parbox[r]{2.9truein}{\epsfxsize=2.9truein\epsfbox{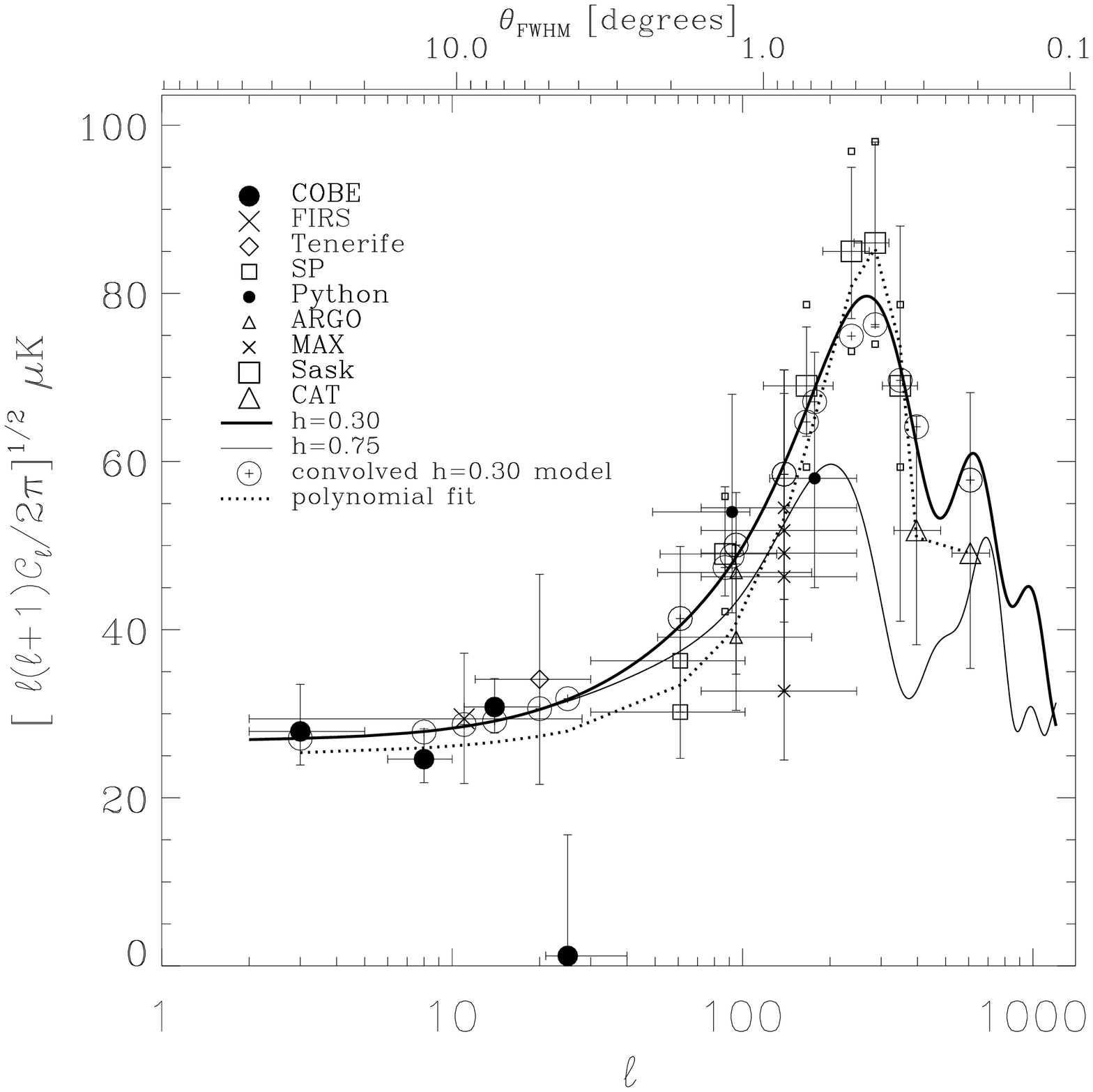}}
\smallskip
}
}

\def\thesecondfig{
\makebox{
\medskip
\noindent
\parbox[l]{1.7truein}{\footnotesize
{\bf Figure 2.  Constraints on Hubble's Constant}
The dark grey areas denote the regions of parameter space favored by the
CMB data. They are defined by $\chi^{2}_{min} + 1$ for 
Sk0 and Sk-14 (minima marked with thick and thin `x' respectively).
`95' denotes the $\chi^{2}_{min} + 4$ contours for
Sk0 (thick) and Sk-14 (thin).
The light grey band is from big bang nucleosynthesis ($ 0.010 < \Omega{b}\:h^{2} < 0.026$). 
The parameters $n$ and $Q$ have been marginalized.
In the $H_{o}$ result quoted, we neglect the region
at $H_{o} \sim 100$ with $\Omega_{b} \sim 0.15$.
This figure shows clearly that lowering the calibration by
14\% {\it does not} favor higher values of $H_{o}$ 
(Figure from Lineweaver \etal 1997b).}
\hglue0.1truein
\parbox[r]{2.9truein}{\epsfxsize=2.9truein\epsfbox{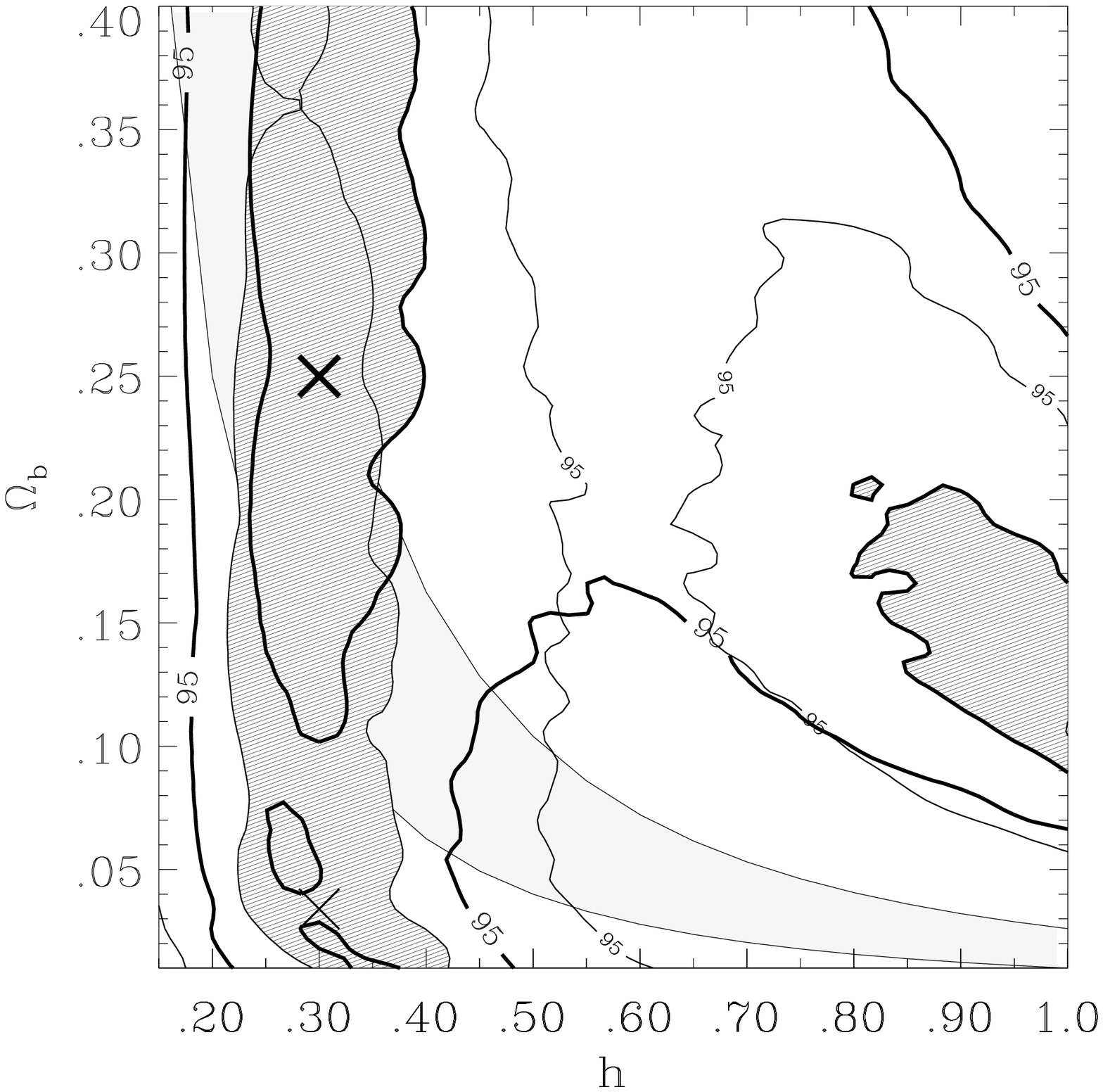}}
\smallskip
}
}

\begin{document}

\title{CONSTRAINING COSMOLOGICAL PARAMETERS\\ WITH CMB MEASUREMENTS}

\author{C.H. LINEWEAVER}

\address{Observatoire Astronomique de Strasbourg\\
         11 rue de l'Universit\'e\\
         67000 Strasbourg, France\\
         charley@astro.u-strasbg.fr}

\maketitle\abstracts{The current enthusiam to measure fluctuations in the CMB power 
spectrum at angular scales between 0.1 and $1^{\circ}$ is largely motivated by the 
expectation that CMB determinations of cosmological parameters will be of 
unprecedented precision.
In such circumstances it is important to estimate what we can already say 
about the cosmological parameters. 
In two recent papers (Lineweaver \etal 1997a \& 1997b)
we have compiled the most recent CMB measurements, used a fast Boltzmann 
code to calculate model power spectra (Seljak \& Zaldarriaga 1996) 
and, with a $\chi^{2}$ analysis, we have compared the data to the power spectra
from several large regions of parameter space.
In the context of the flat models tested we obtain the 
following constraints on cosmological parameters:
$H_{o} = 30 ^{+13}_{-9}$,
$n=0.93^{+0.17}_{-0.16}$ and
$Q=17.5^{+3.5}_{-2.5}\:\mu$K.
The $n$ and $Q$ values are consistent with previous estimates while the 
$H_{o}$ result is surprisingly low.}

\section{Method}

With two new CMB satellites to be launched in the near future 
(MAP $\sim 2001$, Planck Surveyor $\sim 2005$) and half a dozen new 
CMB experiments coming on-line (see contribution of Lyman Page to this volume), 
it is important to keep track of what we can already say about the cosmological parameters. 
In Lineweaver \etal (1997a) we considered
COBE-normalized flat universes with $n=1$ power spectra. We used predominantly
goodness-of-fit statistics to locate the regions of the $H_{o} - \Omega_{b}$
and $H_{o} - \Lambda$ planes preferred by the data.
In Lineweaver \etal (1997b) we obtained $\chi^{2}$ values over the 4-dimensional parameter
space $\chi^{2}(H_{o},\Omega_{b}, n, Q)$ for $\Omega = 1$, $\Lambda = 0$ models.
Projecting and slicing this 4-D matrix gives us the error bars around the minimum $\chi^{2}$
values.
Here we summarize several of our most important results.

\section{Results and Discussion} 

One of the difficulties in this analysis is the 14\% absolute calibration uncertainty of
the 5 important Saskatoon points which span the dominant adiabatic peak in the spectrum
(Figure 1).
We treat this uncertainty by doing the analysis three times: all 5 points at their
nominal values (`Sk0'), with a 14\%
\newpage
\noindent
\thefirstfig
increase (`Sk+14') and a 14\% decrease (`Sk-14').
Sk+14 and Sk-14 are indicated by the small squares in Figure 1 above and below the nominal
Saskatoon points.
Leitch \etal (1997) report a preliminary relative calibration of Jupiter and CAS A 
implying that the Saskatoon calibration should be $-1\% \pm 4\%$.
Reasonable $\chi^{2}$ fits are obtained for Sk0 and Sk-14.

In the context of the flat models tested, our $\chi^{2}$ analysis yields:
$H_{o} = 30 ^{+13}_{-9}$,
$n=0.93^{+0.17}_{-0.16}$ and
$Q=17.5^{+3.5}_{-2.5}\:\mu$K.
The $H_{o}$ result is shown in Figure 2.
The $n$ and $Q$ values are consistent with previous estimates while the 
$H_{o}$ result is surprisingly low.
For each result, the other 3 parameters have been marginalized over.
This $H_{o}$ result has a negligible dependence on the Saskatoon 
calibration, i.e., lowering the Saskatoon calibration from 0 to -14\% 
does not raise the best-fitting $H_{o}$ in flat models.
The inconsistency between this low $H_{o}$ result and $H_{o} \sim 65$ results
will not easily disappear with a lower Saskatoon calibration. 
Our results are valid for the specific models we considered:
$\Omega=1$, CDM dominated, $\Lambda = 0$, Gaussian adiabatic initial
conditions, no tensor modes, no early reionization,
$T_{o}=2.73$ K, $Y_{He}=0.24$, no defects, no HDM.

There are many other cosmological measurements which are consistent 
with such a low
value for $H_{o}$ (Bartlett \etal 1995, Liddle \etal 1996). 
For example, we calculated a joint likelihood based on the 
observations of galaxy cluster baryonic fraction, big bang nucleosynthesis 
and the large scale density fluctuation shape parameter, $\Gamma$. 
We obtained $H_{o}\approx 35^{+6}_{-5}$. 

I am grateful to my collaborators D. Barbosa, A. Blanchard and J. G. Bartlett and 
I acknowledge support from NSF/NATO post-doctoral fellowship 9552722.

\noindent
\thesecondfig
%
%
%
\end{document}